\documentclass{kluwer}    

\usepackage{graphicx}
 
\newdisplay{guess}{Conjecture}
 
\begin{document}
\begin{article}
\begin{opening}
\title{ON THE ABUNDANCE OF HOLMIUM IN THE SUN}
\author{Donald J. \surname{Bord}\thanks{Department of Astronomy, University of Michigan,
Ann Arbor, MI 48109 U.S.A.}}
\runningauthor{Donald J. Bord and Charles R. Cowley}
\runningtitle{On the Abundance of Holmium in the Sun}
\institute{Department of Natural Sciences, University of Michigan-Dearborn,
Dearborn, MI 48128 U.S.A.}
\author{Charles R. \surname{Cowley}}
\institute{Department of Astronomy, University of Michigan, Ann Arbor, MI 48109 U.S.A.}
\date{Submitted \today}
 
\begin{abstract}
The abundance of holmium (\emph{Z} = 67) in the Sun remains uncertain.  The
photospheric abundance, based on lines of Ho II, has been reported as
$+0.26 \pm 0.16$ (on the usual scale where log(H) = 12.00), while the meteoretic
value is $+0.51 \pm 0.02$.  Cowan code calculations have been undertaken to 
improve the partition function for this ion by including important contributions
from unobserved levels arising from the (4f$^{11}$6p + 4f$^{10}$(5d + 6s)$^{2}$)
group.  Based on 6994 computed energy levels, the partition function for Ho II is
67.41 for a temperature of 6000 K.  This is $\approx$1.5 times larger than the value derived
from the 49 published levels.  The new partition function alone leads to an increase
in the solar abundance of Ho to log(Ho) $= +0.43$.  This is
within 0.08 dex of the meteoretic abundance.  Support for this result has been obtained
through LTE spectrum synthesis calculations of a previously unidentified weak line at
$\lambda_{\odot}3416.38$.  Attributing the feature to Ho II,
the observations may be fitted 
with log(Ho) = +0.53.  This 
calculation assumes log$(gf) = 0.25$ and is uncertain by at least 0.1 dex.
\end{abstract}

\end{opening}
 
\section{Introduction}

The abundances of most nonvolatile elements in CI chondrites are in close agreement with
those determined for the solar photosphere (Grevesse and Sauval, 1998; hereafter GS).  The
lanthanide holmium is one of only a few elements with differences of the order of 0.3 dex or more.  GS
give log(Ho$_{\mathrm{CI}}$)$-$log(Ho$_{\odot}$) $= +0.25$.

The solar abundance of holmium reported by GS is $+0.26 \pm 0.16$ and is based on an
unpublished analysis of Ho II, the dominant ion in the Sun, by Daems, Bi\'{e}mont,
and Grevesse (hereafter DBG) in 1984.  Four lines at $\lambda$3343.6, 3398.9, 3456.0, and 
3474.2 {\AA} were analyzed.  All of the features are blended, with the holmium lines constituting 
only minor
contributors to the observed spectrum (N. Grevesse, private communication).  The 
scatter about the mean abundance in this study arises partly from problems of blending and
partly from the uncertainty in the adopted $gf$-values from Gorshkov and Komarovskii (1979). The
partition function for Ho II was calculated using the known atomic levels at the time;
at 6000 K, the value was found to be 43.8 (N. Grevesse, private communciation).

The published level structure for Ho II is seriously incomplete (cf. Martin, Zalubas,
and Hagan, 1978; hereafter MZH): only 49 levels are known, many with no term designations
and a few with multiple J-values.  Wyart, Koot, and
van Kleef (1974) note that a satisfactory analysis of this ion requires the calculation
of the (4f$^{11}$6p + 4f$^{10}$(5d + 6s)$^{2}$) group.  Of particular importance are the
contributions from the 4f$^{10}$6s$^2$ and 4f$^{10}$5d6s configurations which are expected to begin below 12 000
cm$^{-1}$ (Brewer, 1971).  Only eight measured levels have energies lying below this threshold.

Much of the difference between the meteoritic abundance of holmium and that
found by DBG in the Sun may be due to an underestimate of the partition function for Ho II.
Cowan-code calculations were therefore
undertaken to supplement the published data.  In addition, a fresh examination of the solar
spectrum was made in an attempt to identify additional, weak, \emph{unblended} features that might be 
attributable to Ho II and used in spectrum synthesis calculations to establish an independent
estimate of the abundance of this element.  One line in particular, at 3416.38 {\AA,} was found
to be amenable to such analysis.

In the following section, we describe the energy level calculations and give the partition functions that
result from them for temperatures in the range 3000 to 34 000 K.  The third section addresses our
search for other Ho II lines suitable for analysis by spectrum synthesis techniques, while in the fourth,
the results of our study of the $\lambda$3416 line are presented.  The final section of the paper briefly summarizes
the current state of our knowledge of the abundances of the lanthanide rare-earth elements in the Sun
vis-a-vis the CI chondrites; it also includes some remarks about our recent re-examination of the abundance of the
volatile element indium in the Sun.

\section{Energy Level Calculations and Partition Functions}

The methodology employed in making the energy level calculations is based on the Cowan code (Cowan,
1981; 1995) and follows that described by Bord (2000, and references cited therein).  Briefly,
single-particle radial wavefunctions were determined using a Hartree plus statistical exchange
interaction approximation for the following even and \emph{odd} parity configurations, respectively:
(4f$^{11}$6p + 4f$^{10}$(5d + 6s)$^{2}$) and ($4f^{11}6s + 4f^{11}5d$). 
Experimental energy levels exist for three of these configurations, viz. 4f$^{11}$6p, $4f^{11}6s$,
and $4f^{11}5d$, but none is complete.  Relativistic and electron correlation corrections have been included
in the calculations, and the eigenvectors were constructed using both LS- and jj-coupling basis sets.
J$_1$j-coupling is best suited to describe the electron interactions in this ion.

\emph{Ab initio} values for the single-configuration center-of-gravity energies and the various radial
and configuration interaction integrals were computed with uniform scaling applied to all spin-orbit
and Slater parameters.  Reducing the theoretical values of F$^{\emph{k}}$ and G$^{\emph{k}}$ in
this manner roughly accounts for the effects of introducing additional two-body electrostatic operators
for legal values of \emph{k}.  The adopted scale factor for the Slater parameters was 0.68, while that
for the spin-orbit interaction was 0.96.  These values are typical of what has been used in our previous
studies of lanthanide rare-earth ions 
and comport well with values published by other workers
applying the Cowan code in the analysis of these species (cf. Zhang, \emph{et al.}, 2002, and references cited
therein).

The radial integrals for the odd levels were optimized to fit the known energy levels using the method
of least squares.  A total of 468 levels were computed, of which 19 (all that are known) were fitted,
including the ground state.  In as much as the number of known levels equalled the number of structure
parameters to be determined, several of the latter had to be held fixed and/or linked together in order
to secure convergence of the least squares procedure.  After considerable experimentation with various
combinations, the final fit with 9 free parameters yielded a mean deviation in energy of 156 cm$^{-1}$
or $\approx$0.8\% over a range of 20 000 cm$^{-1}$.

Among the even configurations, 6526 levels were calculated.  Since only 11 levels 
possessing reliable energies have been completely
classified, all belonging to the 4f$^{11}$6p
configuration (which requires 9 parameters alone for its description), 
no attempt was made to refine the structure parameters by fitting the measured energies.
Instead, the center-of-gravity energies for each configuration were shifted to produce
agreement between the calculations and the known (in the case of 4f$^{11}$6p) or
predicted (for 4f$^{10}$5d$^{2}$, 4f$^{10}$5d6s, and 4f$^{10}$6s$^{2}$)
energies for the lowest lying levels in each.  In the latter configurations, the estimates of Brewer (1971)
were used.  

Comparisons between the post-shifted calculations and the measured energies in the 4f$^{11}$6p
configuration revealed differences of $\leq500$ cm$^{-1}$ over a range of 30 000 cm$^{-1}$ (or 
$\approx$2\%).  This is typical, in our experience, of the agreement between the raw 
(i.e., un-fitted) calculations and the
experimental energies in rare-earth ions and suggests that similar ``agreement" may be expected for the
low-lying levels in the unobserved (4f$^{10}$(5d + 6s)$^{2}$) group.
Adopting 500 cm$^{-1}$ as the uncertainty
yields an estimated acuracy of $\approx$5\% for the 4f$^{10}$6s$^{2}$ 
and 4f$^{10}$5d6s configurations that begin at about 10 000 cm$^{-1}$ and 11 500 cm$^{-1}$,
respectively.

Partition functions for the range 3000 K to 34 000 K were calculated based on the 6994 Cowan-code levels
following the methodology of Radziemski and Mack (1980) and Cowley and Barisciano (1994).  Where an
unambiguous assignment based on the MZH compilation could be made, the measured energies were substituted
for the calculated values.  Table I shows the results of our computations compared to those that would be
found using only the known/published energies.  As may be seen, at 6000 K, the new partition function is
more than 50\% larger than that found from the levels tabulated in MZH and used by DBG.  This difference
is primarily due to a nearly 12-fold increase in the number of energy levels in the range 10 000 to 30 000
cm$^{-1}$ over what appears in MZH.  As Grevesse (1984) has pointed out, these levels are especially important
for determining the partition functions for ions in the Sun.

\begin{table}
\caption{Partition functions for Ho II}
\begin{tabular}{cccccc}
Temp & This & NIST &
Temp & This & NIST \\ 
(K) & Work & Levels & (K) & Work & Levels \\
 & & & & & \\ 
3000  & 31.09 & 30.31 & 19~000 &1651.95 & 154.08 \\ 
4000 & 37.90 & 34.09 & 20~000 & 1902.26 & 163.40 \\ 
5000 & 49.39 & 38.51 & 21~000 & 2169.26 & 172.59  \\
6000 & 67.41 & 43.63 & 22~000 & 2451.98 & 181.64 \\ 
7000 & 93.93 & 49.50 & 23~000 & 2749.42 & 190.52 \\ 
8000 & 131.02 & 56.09 & 24~000 & 3060.54 & 199.23 \\ 
9000 & 180.77 & 63.35 & 25~000 & 3384.30 & 207.75 \\ 
10~000 & 245.11 & 71.21 & 26~000 & 3719.68 & 216.09 \\ 
11~000 & 325.76 & 79.57 & 27~000 & 4065.66 & 224.24 \\ 
12~000 & 424.08 & 88.34 & 28~000 & 4421.26 & 232.19 \\ 
13~000 & 541.11 & 97.41 & 29~000 & 4785.55 & 239.95 \\ 
14~000 & 677.53 & 106.70 & 30~000 & 5157.61 & 247.51 \\ 
15~000 & 833.65 & 116.13 & 31~000 & 5536.60 & 254.89 \\ 
16~000 & 1009.51 & 125.64 & 32~000 & 5921.68 & 262.07 \\ 
17~000 & 1204.86 & 135.16 & 33~000 & 6312.10 & 269.08 \\ 
18~000 & 1419.22 & 144.66 & 34~000 & 6707.13 & 275.90 \\ 
\end{tabular}
\end{table}

To estimate the sensitivity of the partition functions to uncertainties in the new, calculated
energies, all of which have values greater than 10 000 cm$^{-1}$, 
several numerical tests were performed.  In one series of experiments, the energies
were \emph{randomly} incremented and decremented first by 500 cm$^{-1}$ and then 1000 cm$^{-1}$ and the partition functions
recalculated using the shifted energy values.  At 6000 K, the change in the partition function was
$\leq1\%$ in each case. 
Systematically adding 500 cm$^{-1}$
to every calculated energy level produces a partition function 4.6\% smaller at 6000 K than that
given in Table I; subtracting the same amount from each calculated energy raises
the partition function at this temperature by 5.2\%.  Thus, at temperatures relevant for studies of the Sun,
we conclude that the expected uncertainties in the calculated energies lead to errors in the partition
function of $\approx$5\% or less.  This source thus contributes an uncertainty of $\leq0.02$ dex to the abundance
determination of holmium in the Sun (see below).

The results presented in Table I also comport favorably with the calculations done by Cowley (1984)
based on approximations to the level structure using skewed Gaussians to correct for incompleteness.
Although this approach is limited in its usefulness at relatively low temperatures like that of the Sun, a comparison
between Cowley's partition function for Ho II at 5000 K with that interpolated from Table I reveals
the present Cowan-code--based value to be only $\approx$7\% higher. 
 
Table II gives the empirical fit to the interpolation formula for the partition function
adopted by Bolton (1970), who used a
polynomial in $\mathrm{ln}\, \theta$ (where $\theta = 5040/T$):

\begin{equation}
\mathrm{ln}(u - g_0) = \sum_{i=0}^{n} a_i(\mathrm{ln}\, \theta)^i.
\end{equation}

Here $u$ is the partition function and $g_0$ is the statistical weight of the ground level.  The fits
to the energy sums with $n = 4$ are within 1\% or less for every temperature given in Table I.  Investigators who
prefer to use their own interpolation formulae may obtain the energy levels in machine-readable
form from the first author at \texttt{bord@astro.lsa.umich.edu}.

\begin{table}
\caption{Parameters for polynomial fit to the partition function for Ho II}
\begin{tabular}{cccccc}
$g_0$ &
$a_0$ &
$a_1$ &
$a_2$ &
$a_3$ &
$a_4$ \\
 & & & & & \\ 
17.0 & 3.491604 & -2.318539 & 1.084289 & 0.511300 & 0.036714 \\ 
\end{tabular}
\end{table}

Insofar as the relative number of absorbers scales inversely with the partition function, the larger
the partition function, the greater the required abundance of the species needed to match the
observed line strength.  At the temperature of the Sun (5770 K), increasing the partition function by a factor
of 1.48 forces an increase in the holmium abundance of 0.17 dex.  This yields a value of log(Ho) =
+0.43, based on the earlier results (DBG).  Even with no other changes, this brings the abundance of 
holmium within 0.08 dex of the meteoritic value.  The estimated uncertainties in the computed energies
and, hence, in the new partition function add negligibly to the errors quoted by DBG ($\pm0.16$ dex).

\section{Holmium Line Identifications}

Moore, Minnaert, and Houtgast (1966; hereafter MMH) include no lines of Ho II in their
count of individual spectra in the Sun.  Grevesse and Blanquet (1969) searched for the
strongest Ho II lines in the solar spectrum and found that all but seven were masked by
well-identified lines, including three of those ultimately used by DBG.  
Because of differences between the laboratory wavelengths of the seven surviving lines and those of
unidentified solar lines ranging from 0.013 to 0.083 \AA, these authors concluded that it was
premature to associate any solar feature entirely with Ho II.

Only four of the seven Grevesse-Blanquet (GB) candidate lines have been classified and have published $gf$-values.
We have carefully re-examined the solar spectrum in the vicinity of where these lines might be found.  Our
purpose was to see if a compelling case could be made for an identification that could potentially produce
an accurate holmium abundance by spectrum synthesis.  Our survey revealed no good candidates from among the four
lines considered. We briefly note the results of our investigations below: 

$\mathbf{\lambda4045.4}$:  This line was suggested by GB for association with a 9 m\AA~feature 
reported by MMH at
$\lambda_\odot4045.508$.  No local minimum was discernible near this
wavelength in high resolution solar spectra obtained at Kitt Peak (Neckel, 1999) or the Jungfraujoch
station (Delbouille and Roland, 1995).  The putative feature is likely hidden in the wing of the very strong Fe I
line at $\lambda_\odot4045.825$.  This Ho line has been used, however, in the determination of
the holmium abundance in the metal-poor halo giant $\rm{CS}\,22892-052$ by Sneden, \emph{et al.} (1996).

$\mathbf{\lambda3796.7}$:  This line, arising from the ground level and having log($gf$) $= 0.20$,
has also been used in the analysis of $\rm{CS}\,22892-052$.  In the Sun, an unidentified
33 m\AA~feature at $\lambda_\odot3796.803$
appears much too strong to represent a reliable candidate for
a holmium identification.

$\mathbf{\lambda3474.2}$:  This line was suggested by GB for identification with a 5.5 m\AA~feature
at $\lambda_\odot3474.273$, and was included in the analyses of DBG.  This feature is a blend of
two Fe I lines ($\lambda\lambda3474.26$ and 3474.27), a line of Dy II ($\lambda3474.27$),
and the holmium line, with the latter being a distinctly minority contributor to the feature.  Indeed,
we were able to satisfactorily fit the observed solar feature by spectrum synthesis by including
only the dysprosium and iron lines at their solar abundances.  Adding the holmium
line at the CI abundance produced no discernible change in the calculated profile; this occurs in
large part because the hfs of this feature shifts the strongest component of the red-degraded flag
pattern to 3474.160 \AA~in the wing of a very strong Mn II line. Although our results are thus
compatible with a holmium abundance equal to that found in the meteorites, the insensitivity of 
this feature to variations in the holmium abundance makes it unsuitable for
a definitive test of the amount of this element in the Sun.

$\mathbf{\lambda3416.4}$:  This line will be discussed in the next section.

A fifth Ho II line at $\lambda4152.59$, not included by GB, was also
considered by us as a candidate for analysis.  This line has also been used by Sneden, \emph{et al.} (1996)
in their analysis of $\rm{CS}\,22892-052$, and was posited by us for association with an unidentified 
8 m\AA~line at $\lambda_\odot4152.527$ in the Sun (MMH).
We computed the hfs splitting of the
levels involved in the transition producing this line, taking the A and B parameters for the
lower level from Worm, Shi, and Poulsen (1990) and those for the upper level from Sneden, \emph{et al.} (1996).  The
computed energy shifts (in cm$^{-1}$) were applied to the ``mean position" (also in cm$^{-1}$) of the
line to produce the hfs pattern.  The appropriate ``mean position" to use was not
obvious, however.  Monograph 145 (Meggers, Corliss, and Scribner, 1975) gives $\lambda_{lab} = 4152.61$ \AA;
Sugar (1968) reports a wavelength of 4152.62 \AA, and, based on the energy levels in MZH, we
compute a wavelength of 4152.59 \AA.  Sneden, \emph{et al.} (1996) adopted 4152.58 \AA~in their study.
Nave (2001, private communication) has recently measured this feature on a high resolution FTS spectrum and gives
the center-of-gravity wavelength as 4152.604 \AA.  Using this value, we find the strongest hf
component to lie at 4152.441 \AA, in excellent agreement with Nave's hfs measurements.  These calculations
place the bulk of the holmium absorption near 4152.45 \AA~at a local \emph{maximum} in the solar spectrum
and far from the observed $\lambda_\odot4152.53$ feature.  In light of this discordance, 
further exploration of this line was abandoned
in favor of the more promising feature near 3416.4 \AA.

\section{Spectrum Synthesis of the $\lambda3416.4$ Feature}

The $\lambda3416$ line in Ho II arises from the following transition:
$637.40 (J=7) - 29 899.21 (J=7)$.  Precision FTS data put the center-of-gravity wavelength at
3416.444 \AA~(G. Nave, private communication).  The hfs constants A and B for both the upper
and lower levels of this transition are given by Worm, Shi, and Poulsen (1990) and yield a
wavelength of 3416.376 \AA~for the strongest component of the primary octet pattern.  This
value compares very favorably with the position of a feature at 3416.375 \AA~in the
Jungfraujoch solar spectrum for which we measure an equivalent width of 6.4 m\AA.  MMH report a feature
with $\lambda_\odot
= 3416.409$ \AA~and $W_\lambda = 6$ m\AA, but give no identification.
We adopt log($gf$)$ = 0.25 \pm 0.04$ based on Nave's transition
probability; this is in good agreement with the VALD value of 0.23 for this transition (Kupka, 
\emph{et. al.}, 1999).

A 2.25 \AA~segment centered on $\lambda3416.4$ was synthesized in LTE using the suite of programs
described by Cowley (1996) and compared to both the Jungfraujoch and Kitt Peak solar spectra.  
The observed spectra are practically indistinguishable in the vicinity of the $\lambda3416$
feature, and both yield a relative minimum in the continuum at $3416.375 \pm 0.001$ \AA.

Line data for atomic
species (excepting holmium) were extracted from VALD, and the molecular data were taken from
Kurucz (1993).  Observed and predicted features for the (0,0), (1,1), and (2,2) bands of NH were
included in the calculations, although only the transitions with the lowest J-values for the
(2,2) system were used; the higher series members are weak and/or absent due to pre-dissociation
at the temperature of the Sun.
Small modifications to the Kurucz wavelengths for NH, typically amounting to less than
$0.005$ \AA,  were made based on the
energy levels published by Brazier, Ram, and Bernath (1986).  Solar abundances (GS) of all atomic
species save holmium were adopted, and the NH abundance was reduced by a factor of 0.55
from its equilibrium value derived using the solar abundances of N and H in order to
achieve acceptable fits to the observed lines.

A 137-step model atmosphere based on the 
$T(\tau)$ given by Holweger and M\"{u}ller (1974) was employed in this study.  
A 137-step version of a Kurucz model
atmosphere with $l/h = 0.75$ and no convective overshoot (R. Kurucz, private comunication) was also tested for comparison.  
It predicted a slightly higher continuum level in the vicinity of the $\lambda3416$ feature,
which led to the need to reduce the optimum holmium abundance (see below) by 0.1 dex in order to 
reconcile the calculations to the observations.  This suggests that uncertainties
in the model atmosphere and the adopted atomic parameters produce errors in the final
solar holmium abundance of between 0.10 and 0.15 dex overall.

The spectrum was calculated in units of the specific intensity in double precision with the
following model parameters: $v sin i = 0.0$ km s$^{-1}$  and $ v_{turb} = 1.0$ km s$^{-1}$.  The 
computed spectrum was broadened using a Gaussian with half width 0.018 \AA~to match the
resolution of the observations.  The continuous opacity was also increased by a factor of 1.2
to achieve agreement between the calculated emergent intensity and the measurements of
Neckel and Labs (1984) at 3400 \AA.

Figure 1 compares the spectrum synthesis (solid line) with the observations (dotted line)
\emph{assuming no holmium is present in the Sun.}  The absence of holmium in the model
calculations makes it impossible to fit the observations near 3416.4 \AA~despite the generally
good agreement between the synthesized spectrum and the solar data throughout most of the
remainder of the region.  (A notable exception is the unidentified feature at $\lambda_{\odot}3416.51$
discussed below.)  Clearly, some opacity due to this element is required in order
to reconcile the calculations with the observations.  The issue remains, how much holmium
is needed.

\begin{figure}
  \centerline{\resizebox{12cm}{!}{\includegraphics*[30mm,2mm][280mm,216mm]{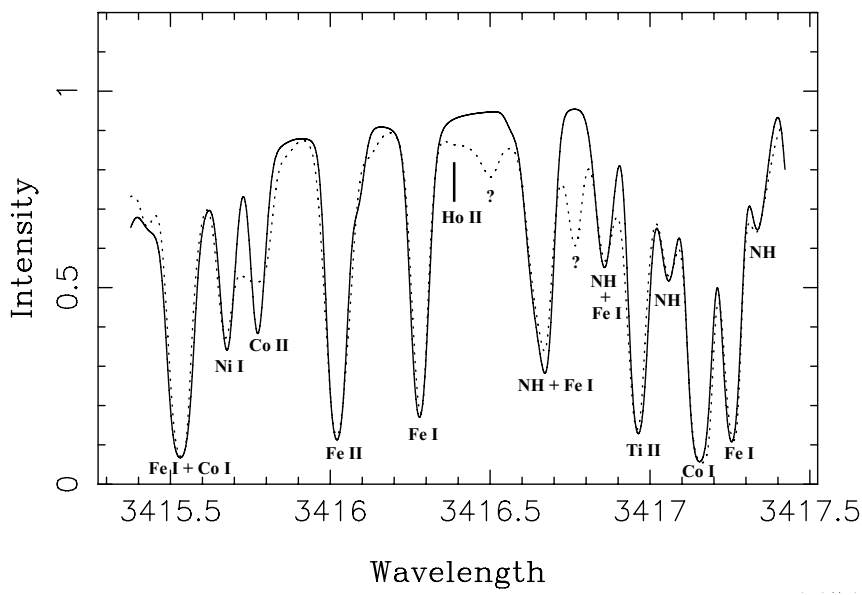}}}
  \caption{Synthesis (solid line) of the $\lambda3416.4$ region
of the solar spectrum {\bf with no absorption due to holmium included in the model.}
The observed Jungfraujoch spectrum is indicated with the dotted line and is not fit
at all by the computations in the vicinity of the putative Ho II feature at 3416.38 \AA.
Remaining identifications are from MMH; absorptions of unknown origin are indicated
by question marks.  No attempt has been made to
improve the overall fit by including the hfs of the cobalt lines, although some adjustments
to the
wavelengths and $gf$-values of a few lines were made to achieve better agreement with the
observations as described in the text.}
\end{figure}

It should be noted that in Figure 1, no attempt has been made to improve the fit by
including the hfs of the cobalt lines appearing in the region, although some adjustments
to the VALD $gf$-values of a few lines (e.g. Ni I $\lambda3415.68$ and Fe II $\lambda3416.09$)
were made to achieve better agreement with the observations.  Also, because of its potential
influence on the structure of the spectrum immediately to the blue of the Ho II feature at
3416.38 \AA, care was taken to fit the Fe I line at 3416.28 \AA~as well as possible.  In
particular, small adjustments have been made in the VALD-values of the log($gf$) (reduced by 0.2 dex),
the wavelength (reduced by 0.003 \AA), and the damping width (increased by 20\%).
As it turned out, when tests were made, it was discovered that the best-fit holmium abundance
was insensitive at the level of
less than 0.05 dex to the modest modifications of the VALD values proposed above.  In the
figures that follow, the calculations have incorporated the modified choices for these parameters.

MMH also identify a feature at
$\lambda_{\odot}3416.512$ as a predicted line of Fe I, but no line with this
approximate wavelength appears in the VALD database or in the recent work on this ion
by Nave, \emph{et al.} (1994). Synthesis of this feature using only weak lines of Cr I
and Ti I in the VALD list at $\lambda\lambda3416.501$ and $3416.504$, respectively,
clearly produces poor agreement with the observations (cf. Figure 1),
confirming some degree of missing line opacity in this region.  In the absence of complete data,
an attempt was made to better
fit this absorption by making small adjustments to the wavelengths of the Cr I and Ti I
lines and by increasing their oscillator strengths by about 3 dex each. The resultant profile is still not
a completely satisfactory fit to the observations, as may be seen in subsequent figures, but it
provides a reasonable match to the red wing of the Ho II feature and serves to demonstrate the
influence of the extended hfs of this line. 

\begin{figure}
  \centering
  \resizebox{\textwidth}{!}{\includegraphics*[0mm,2mm][240mm,216mm]{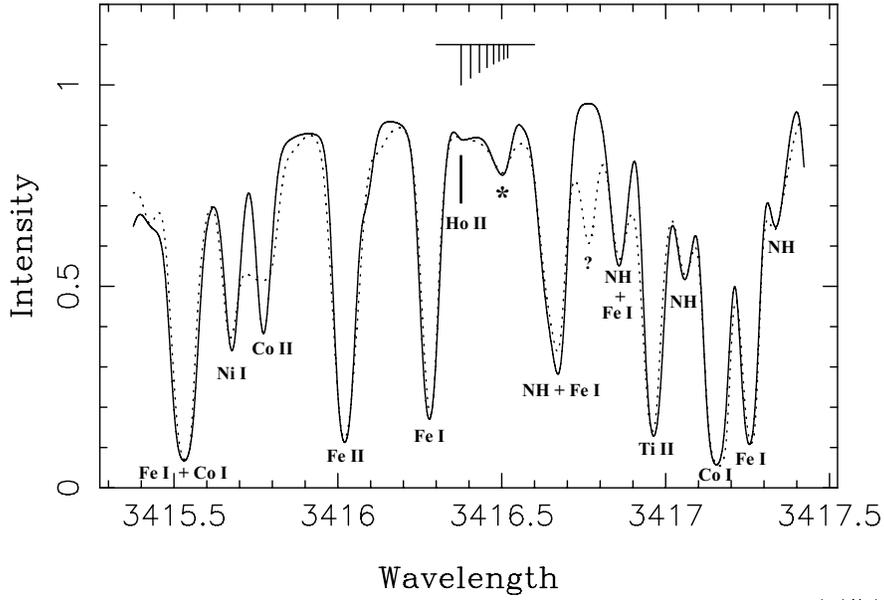}}
  \caption{Synthesis (solid line) of the $\lambda3416.4$ region
of the solar spectrum with an adopted holmium abundance of +0.53 on the usual scale where log(H) = 12.00.
The observations are again given by the dotted curve.  The primary octet pattern in the hfs for the
Ho II feature is also shown.  The fit also reflects modifications in the positions and
oscillator strengths of Ti I and Cr I lines near 3416.5 \AA~(asterisk symbol) as described in the text to better
model the unidentified absorption (cf. Figure 1) immediately to the red of the Ho II line of interest.
All other identifications follow Figure 1.}
\end{figure}

Figure 2 shows our synthesis (solid line) of this region for log(Ho) = +0.53.  The observations
are again given by the dotted curve.  The primary hyperfine pattern for the Ho II
feature is shown, with intensities drawn approximately to scale.  Although only the
eight principal lines are given in the figure, all twenty-two have been included in the
synthesis.  As isotope 165 comprises 100\% of naturally occurring holmium, no additional
wavelength shifts due to isotope effects are neccessary.  Gravitational redshift effects on
the wavelength of this line have been ignored in the calculation.  As may be seen, with this choice
of holmium abundance, the weak solar feature at 3416.375 \AA~is accounted for in a respectable
fashion.  This figure also shows the effects of ``mocking up" the unidentified absorption at
$\lambda_{\odot}3416.51$ by adjusting the VALD parameters for the Cr I and Ti I
lines as remarked upon above.

\begin{figure}
  \resizebox{\textwidth}{!}{\rotatebox{270}{\includegraphics{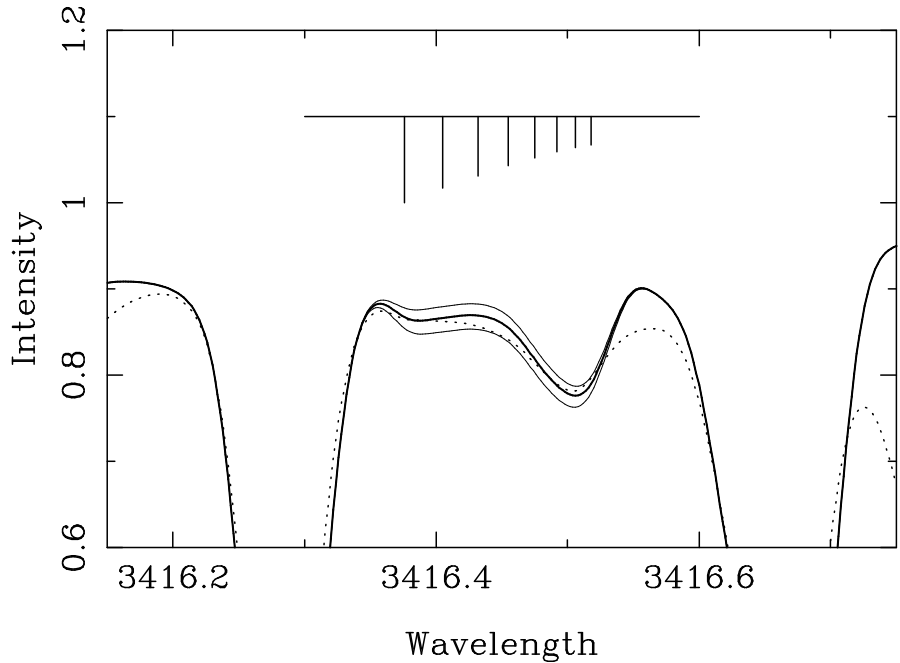}}}
  \caption{Detail of solar spectrum in the vicinity of the Ho II 3416.44 \AA~line.
As in Figure 2, the heavy solid line is the computed spectrum for log(Ho)=+0.53, while the
observed spectrum is dotted.  The upper(lower) thin lines show synthetic calculations with the
holmium abundance decreased(increased) by 0.1 dex.  The fit is obviously poorer near the
positions of the strongest hfs components for the augmented(decremented) spectrum.  The primary hfs
for the Ho II line is again shown for reference, and its
influence of the structure of the unidentified absorption feature at 3416.51 \AA~is obvious.}
\end{figure}

Figure 3 highlights the wavelength region in the immediate vicinity of the Ho II 
line and shows the effects of small variations in the abundance of this element.  In particular,
the thin solid lines show the calculated spectrum with the holmium abundance increased (bottom
curve) or decreased (upper curve) by 0.1 dex with respect to the value adopted in Figure 2. 
As may be seen, the fit is demonstrably poorer near the positions of the strongest
hfs components for the augmented and decremented
plots.  By additional fine tuning, a slightly better fit to the observations than that
afforded by the heavy solid line might be achieved, 
but, given our earlier comments
regarding the formal errors in the computations attributable to uncertainties in the adopted
model atmosphere and the atomic parameters, we did not see much profit in doing so.  As it is, our
nominal best-fit value for the solar holmium abundance falls within 0.02 dex of the meteoritic value
reported by GS, 
while prior considerations persuade us
that our result is accurate to $\pm0.15$ dex at best.  We are reassured, to some degree
however, of the
validity of our result by the agreement between the
holmium abundance derived from our best fit to the solar spectrum and that found by simply
correcting the estimate of DBG to account for improvements in the partition function for
this ion.  

\section{Conclusions}

Uncertainties in the spectrum synthesis not withstanding, the results presented herein
lend credence to the view that the abundance of holmium in the Sun follows the
pattern of other rare-earth elements in showing general agreement with the meteorites.
In Figure 4 we plot the logarithmic CI abundances from the compilation of McDonough and Sun (1995) 
minus the solar abundances from
GS (with some small modifications to reflect recent work on the light elements by Holweger (2001)). 
Because McDonough and Sun have not been involved with the reconciliation of solar and meteoritic
abundances, they may be considered an independent source for the meteoritic values,
which are, in any case, very close to those reported by GS.  (For example, McDonough and
Sun give the logarithmic abundance of holmium in the meteorites as +0.49, 0.02 dex
lower than the value quoted by GS but within the latter's stated error.)  The abundance differences are
plotted versus elemental condensation temperatures taken from Lodders and Fegley (1998). 

\begin{figure}
  \resizebox{10cm}{!}{\rotatebox{270}{\includegraphics{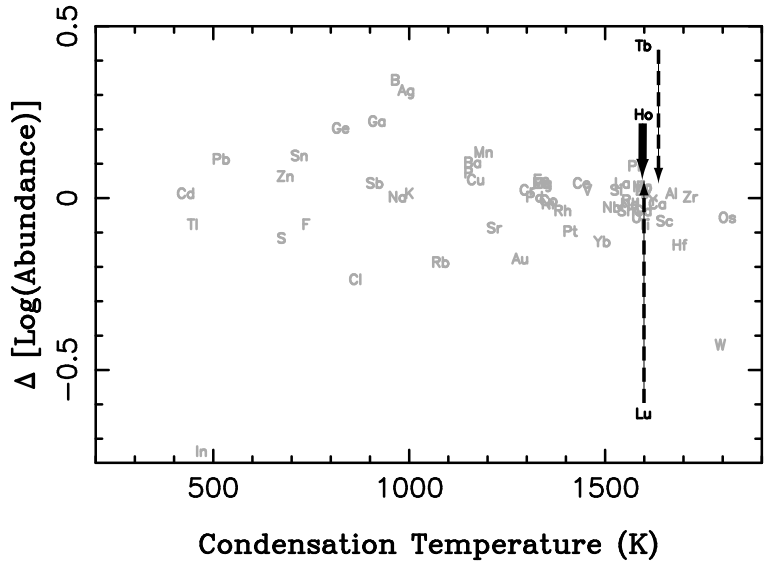}}}
  \vspace{1.5cm}
  \caption{ Logarithmic CI minus solar photospheric abundances vs. elemental
condensation temperatures.  Points are indicated by the chemical symbols of the elements.
The position of holmium prior to and after the present work is shown by the heavy arrow.
Recent adjustments in the abundances of lutetium and terbium
are shown by the dashed arrows.  Among the heavy and
rare-earth elements, only tungsten~(W) remains seriously discordant.}
\end{figure}

As may be seen in Figure 4, with recent revisions to the solar abundances of lutetium (Bord, Cowley, and
Mirijanian, 1998), terbium (Lawler, \emph{et al.}, 2001) and now holmium, 
among the heavy and rare-earth elements, only tungsten remains
seriously anomalous.  The authors are currently investigating the prospects of being
able to improve the photospheric abundance of this element using the techniques described in
this paper. 

Among the volatile elements, indium ($Z = 49$) presents the greatest discordance.  We have recently re-evaluated
the solar abundance of this element in a manner similar to that described in this paper (cf. Bord and
Cowley, 2001).  Specifically,
we synthesized a 2.25 \AA~segment of the solar spectrum centered on the position of the resonance
line of In I at $\lambda4511.3$.  The hfs constants were taken from Jackson (1957, 1958), and the
VALD log($gf$) = $-0.210$ was adopted.  Our best fit to the solar data yielded log(In) = 1.56.  This is about
0.1 dex smaller than the value found by Lambert, Mallia, and Warner (1969), the difference arising almost
entirely from revisions in the oscillator strength of the line.  We estimate that the formal errors in the 
computation are at least $\pm0.2$ dex, half of which is due to residual uncertainties in the oscillator
strength.  The remainder of the error is contributed by uncertainties associated with the placement of
the continuum which amount to about 1\% in this region.  We have thus confirmed that the Sun is overabundant
in indium by a factor of more than 5, with log(In$_{\mathrm{CI}}$)$-$log(In$_{\odot}$) $= -0.74$.  

This
difference is reflected in Figure 4 and remains a challenge to our understanding.  With no other
unblended indium lines available, further refinements in the solar abundance of this element appear
remote.  Moreover, the meteoritic abundance rests on 24 separate analyses made at three different labs
and is uncertain by only 6\% (Anders and Ebihara, 1982).  Given that indium should be one of the last
of the chalcophiles to leave the gas phase, conditions in the solar nebula may have led to its incomplete condensation
and to a depletion of this species in the CI meteorites.  If this were to be the case,
one might expect to see a similar depletion of cadmium ($Z = 48$) in these meteorites relative to the
Sun.  However, the cadmium abundance in the CI meteorites is within 0.01 dex of that found for the solar
photosphere (Youssef, D\"{o}nszelmann, and Grevesse, 1990), so the mystery persists.  

These exceptions not withstanding, 
it is quite remarkable - indeed, puzzling perhaps -  the degree to which the abundances of the CI meteorites
agree with those of the solar photosphere.  Given the complex chemical histories of these meteorites
in which aqueous alteration has likely played a significant role, it is surprising that the bulk
atomic compositions of these bodies seem to have remained largely unfractionated.  Continuing efforts to
refine the abundances of trace elements in the Sun to permit more precise comparisons
between the abundance patterns in our star and these meteorites hold promise for providing greater
insight into the history of both. 

\acknowledgements{The degree to which the spectrum synthesis calculations undertaken herein are secure
depends critically on the accuracy of the atomic parameters for Ho II.  The
results reported here have benefited significantly from improvements in the accuracy
of wavelengths, energy levels and oscillator strengths arising from high resolution
spectroscopic studies of this ion being carried out at NIST and at the University
of Lund.  It is a pleasure to acknowledge helpful communications with Drs. Gillian
Nave (NIST) and Glenn Wahlgren (Lund) during the course of this investigation on this
subject.

Special thanks are also due to our colleagues Nicolas Grevesse (Li\'{e}ge) and Jacques Sauval
(Royal Observatory, Belgium) for help
and advice on matters pertaining to solar abundances, and to Robert Kurucz (CfA) for data and counsel
concerning solar model atmospheres.  We are most grateful for the continuing guidance of 
Robert Cowan (LANL) on the use of his codes, as well as on general problems of atomic
spectra and structure.}

\end{article}
 
\end{document}